\documentclass[runningheads]{llncs}
\usepackage{hyperref}
\usepackage{caption}
\usepackage{gensymb,graphics,graphicx,array,amsmath,caption,amsfonts,algorithm,algorithmic}
\usepackage{subfigure,psfrag,epsfig,epstopdf,amssymb,array,flushend}

\begin{document}
\title{A Robust Blind  3-D Mesh Watermarking technique  based on SCS quantization \\ and mesh Saliency  for Copyright Protection}
\author{Mohamed Hamidi\inst{1}
\and
Aladine Chetouani\inst{2} \and
Mohamed El Haziti\inst{1} \and Mohammed El Hassouni\inst{1,3}  \and Hocine Cherifi\inst{4}}
\authorrunning{Mohamed Hamidi et al.}
%
\institute{LRIT - CNRST URAC29, Rabat IT Center, Faculty of Sciences, Mohammed V University in Rabat, Morocco
\email{\{hamidi.medinfo,elhazitim\}@gmail.com}
 \and
PRISME Laboratory, University of Orleans, France
\email{aladine.chetouani@univ-orleans.fr}\\
 \and
LRIT - CNRST URAC29, Rabat IT Center, FLSH ,\\ Mohammed V University in Rabat, Morocco
\email{mohamed.elhassouni@gmail.com} 
 \and
 Le2i, UMR 6306 CNRS, University of Burgundy, France
\email{hocine.cherifi@u-bourgogne.fr}}
%
%
%
%
\maketitle              
\begin{abstract}
Due to the recent demand of 3-D meshes in a wide range of applications such as video games, medical imaging, film special
effect making, computer-aided design (CAD), among others, the necessity of implementing 3-D  mesh watermarking schemes  aiming to protect copyright has increased in the last decade.
 Nowadays, the majority of robust 3-D watermarking approaches have mainly focused on the robustness against attacks while the  imperceptibility of these techniques is still a serious challenge.
  In this context, a blind robust 3-D mesh watermarking method based on mesh saliency and scalar Costa scheme (SCS)  for Copyright protection is proposed. 
The watermark is embedded by quantifying the vertex norms of the 3-D mesh by SCS scheme using the vertex normal norms as synchronizing primitives. The choice of these vertices is based on  3-D mesh saliency to achieve watermark robustness while ensuring high imperceptibility. 
 The experimental results show that in comparison with the alternative methods, the proposed work can achieve a high imperceptibility performance while ensuring a good robustness  against several common  attacks including similarity transformations, noise addition, quantization, smoothing, elements reordering, etc.

\keywords{3-D mesh watermarking   \and scalar Costa scheme (SCS) \and mesh saliency \and Copyright protection.}
\end{abstract}

\section{Introduction}
Nowadays,  the transfer of multimedia contents such as  image, audio, video and 3-D model has been increased considerably due to  the increase in network bandwidth and the rapid development of digital  services. These contents can be copied or  modified easily. 
Therefore, the necessity to protect their copyright  becomes crucially important. Digital watermarking has been found as an efficient solution to overcome this problem. Its underlying concept is to embed an auxiliary information named watermark  into multimedia content to protect its ownership. It is worth noticing that only 3-D meshes are considered in the proposed work.
	 3-D meshes have been widely used in several applications including medical images, computer aided design (CAD), video games, virtual reality, film special effect making, etc.
	  A 3-D mesh is defined as a collection of polygonal facets that aim to give an approximation of a real 3-D object. It contains three  primitives : vertices, edges and facets. A mesh can be also described by two kinds of information: geometry information that represents the positions of vertices and connectivity information which describes the adjacency relations between the different components.
	
 Three major requirements must be satisfied in each watermarking system : imperceptibility, capacity and robustness. Imperceptibility represents  the similarity between the original 3-D mesh and the watermarked one while the capacity refers to the maximum amount of information that can be embedded in the  3-D mesh. Robustness means the ability of extracting the watermark bits even if the stego model has incurred manipulations named attacks. These attacks  can be divided into two main groups. Connectivity attacks, that include   subdivision, cropping, remeshing,  and simplification. Geometric attacks including local deformation operations, similarity transformations as well as signal processing operations. The applications of 3-D mesh watermarking include authentication, indexation, content enhancement, copyright protection, etc. 

It is worth mentioning that compared to the maturity of image, audio and video watermarking methods \cite{hamidi2018MTAP}, there are only few 3-D mesh watermarking techniques that have been proposed \cite{hamidi2017robust} . This is mainly due to the challenges in three dimensional geometry related to its irregular and complex topology as well as the gravity of attacks that 3-D meshes can be exposed to \cite{Survey2007}. Moreover, in contrast with 2-D image, there is no obvious robust intrinsic ordering for 3-D mesh elements, which make the use of spectral analysis watermarking of 2-D images impossible.

The majority of 3-D watermarking methods have mainly focused on the robustness against attacks while few ones based on saliency have been proposed \cite{hamidi2019blind}. Nakazawa et al. \cite{nakazawa2010visually} presented a 3-D watermarking   method based on visual saliency. First, the perceptually conspicuous regions have been identified using the mesh saliency of Lee et al. \cite{lee2005mesh}. Next, the norm of each vertex is calculated and its histogram is constructed. Finally, the watermark is embedded in each bin by normalizing the associated vertex norms. 
In \cite{zhan2014blind}, Zhan et al.  proposed a blind 3-D mesh watermarking method based on curvature. The authors calculated the root mean square curvature for all vertices and the watermark is embedded by modulating the mean of the root mean square curvature fluctuation of vertices. Rolland-Neviere et al. \cite{rolland2014triangle} proposed a 3-D mesh watermarking method where the watermark embedding is formulated as a quadratic programming problem.  
In \cite{son2017perceptual}, Jeongho Son et al.  proposed a 3-D watermarking technique that aims to  preserve the appearance of the 3-D watermarked  model. The authors used  the distribution of the vertex norm histogram as a watermarking primitive  which has been already introduced by Cho et al. \cite{Cho2007}. The latter embeds the watermark by modifying the mean or variance of the vertex norms histogram. 

In this paper, a 3-D robust and  blind watermarking method based on mesh saliency and scalar Costa scheme (SCS) quantization is proposed. The watermark is embedded in the  3-D mesh by quantizing its vertices norms according to the mesh saliency. The  vertex normal norms are sorted in the descending order and chosen as synchronizing primitives. This order is found  to be robust to geometric attacks and element reordering. Taking the full advantages of mesh saliency as well as SCS scheme, the proposed method can ensure  both high robustness to common attacks and good imperceptibility. The rest of this paper is organized as follows. Section \ref{3-D mesh saliency} presents the 3-D mesh saliency. Section \ref{Proposed method} gives a description of the proposed method composed by embedding and extraction. The experimental setup, evaluation metrics and experimental results are discussed in Section \ref{Experimental results}. Finally, Section \ref{Conclusion} concludes the paper.
\section{3-D mesh saliency}
\label{3-D mesh saliency}
Nowadays, saliency detection becomes an interdisciplinary scientific study of computer science and  human perception. It allows to   detect perceptually important points or regions of a 3-D mesh automatically \cite{song2014mesh}. Mesh saliency can be defined as a  measure that captures the importance of a point or local region of a 3-D mesh in a similar way to human visual perception. 
 This technique generally merges perceptual criteria inspired by  human visual system (HVS) with mathematical measures based on geometry. The visual attention of Human is usually directed to the salient shape of the 3-D model. The evaluation of mesh saliency used  in the proposed scheme is Lee et al. \cite{lee2005mesh}. Lee's method evaluates the saliency of each vertex using the difference in mean curvature of the 3-D mesh surfaces from those at other vertices  in the neighborhood. The first step is computing surface curvatures. The computation of the curvature at each vertex $v$  is performed using Taubin's method \cite{taubin1995estimating}. Let $Curv(v)$ the mean curvature of a mesh at a vertex $v$. The Gaussian-weighted average of the mean curvature can be expressed as follows:  
\begin{equation}
G(Curv(v),\sigma )=\frac{\sum_{x \in N(v,2\sigma)} Curv(x)exp(\frac{-\parallel x-v \parallel^2}{2\sigma^2})}{ \sum_{x \in N(v,2\sigma)}exp(\frac{-\parallel x-v \parallel^2}{2\sigma^2})}
\label{eq:GaussAvergMeanCurvature}
\end{equation}
where $x$ is a mesh point and $N(v,\sigma)$ denotes the neighborhood for a vertex $v$ which represents a set of points within an Euclidean distance $\sigma$ calculated as : 
\begin{equation}
N(v,\sigma)=\{x | \parallel x-v \parallel < \sigma\}
\end{equation}
The saliency $S(v)$ of a vertex $v$ is calculated as the absolute difference between the Gaussian-weighted averages computed at fine and coarse scale. 
\begin{equation}
S(v)=|G(Curv(v),\sigma) - G(Curv(v),2\sigma)|
\end{equation}

\begin{figure}[!h]
\centering
\includegraphics[width=0.35\textwidth]{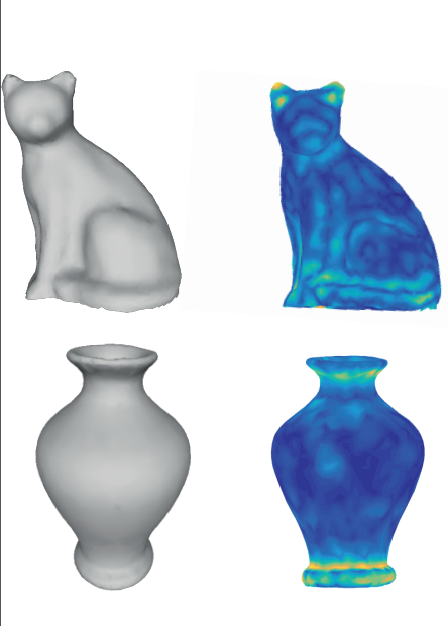}
\caption{3-D meshes and their corresponding mesh saliency  using Lee's method \cite{lee2005mesh} : (left) Original 3-D meshes, (right) 3-D mesh saliency.}
\label{fig:SaliencyVsMeshes}
\end{figure}

Figure \ref{fig:SaliencyVsMeshes} reported an example of mesh saliency of Cat and Vase object using Lee's method \cite{lee2005mesh}.

\section{Proposed scheme}
\label{Proposed method}
\begin{figure}[!h]
\centering
\captionsetup{justification=centering}
\begin{center}
\includegraphics[width=0.70\textwidth]{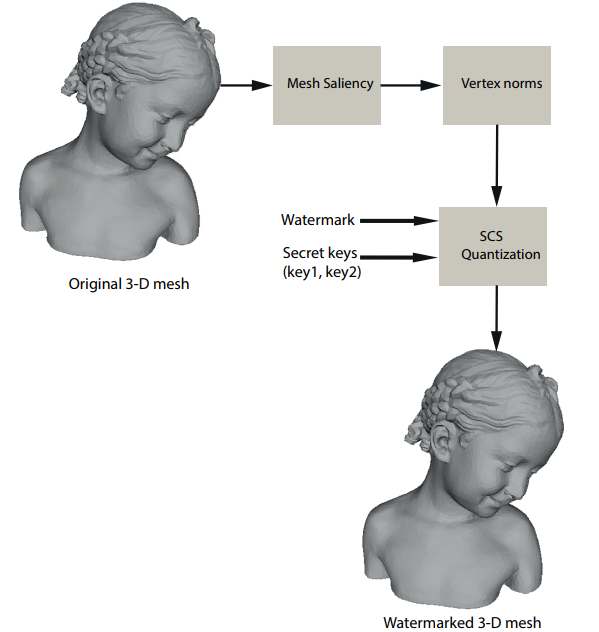}
\end{center}
\caption{The proposed embedding scheme.}
\label{fig:EmbeddingScheme}
\end{figure}

\begin{figure}[!h]
\centering
\captionsetup{justification=centering}
\includegraphics[width=0.72\textwidth]{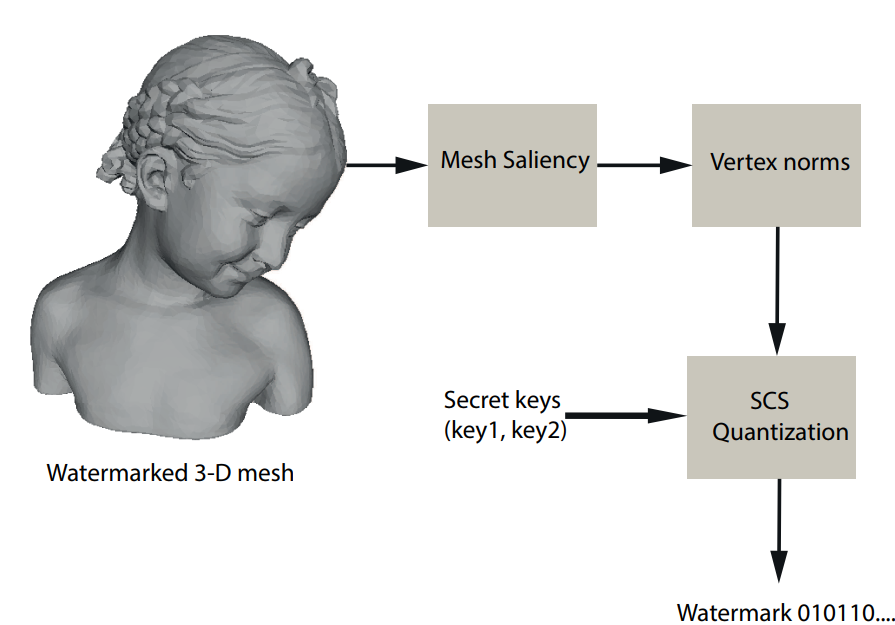}
\caption{The proposed extracting scheme.}
\label{fig:ExtractingScheme}
\end{figure}

\begin{figure}[!h]
\centering
\captionsetup{justification=centering}
\includegraphics[width=0.56\textwidth]{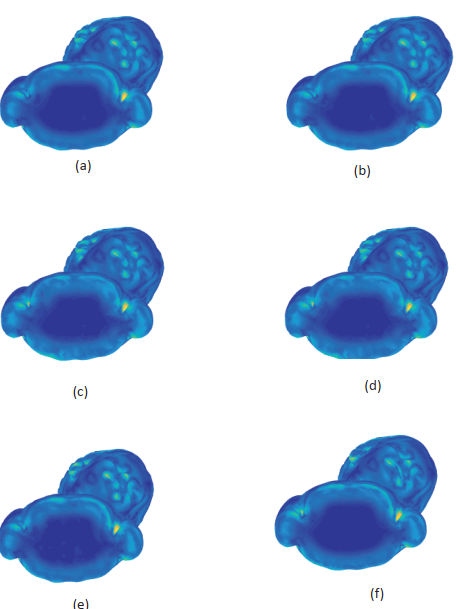}
\caption{Mesh saliency of Bimba before and after attacks : (a)  before attack, (b) additive noise $0.3$\%, (c) Similarity transformation $1$, (d) Simplification ratio, (e) Quantization $9$ bits, (f) Smoothing $\lambda=0.1$  ($30$ iterations).}
\label{fig:StabilitySaliencyAttacks}
\end{figure}

The majority of 3-D  robust watermarking approaches have focused on the resistance to attacks. Only few methods have  investigated the visual impact caused by the watermark embedding. In this context, a robust and blind  3-D mesh watermarking method based on mesh saliency and SCS  quantization for Copyright protection is proposed.  In this work, the watermark is embedded by quantifying the salient vertex norms of the 3-D object using mesh saliency proposed by Lee et al. \cite{lee2005mesh} and SCS quantization. Lee's method is used to obtain candidate vertices in order to ensure high imperceptibility and to improve the robustness performance. The choice of these points is motivated by the fact that these primitives are  relatively stable even after that 3-D meshes suffered from different attacks such as additive noise, similarity transformations, quantization, smoothing, etc. Figure \ref{fig:StabilitySaliencyAttacks} illustrates the  Lee's \cite{lee2005mesh} mesh saliency of Bimba model after applying different attacks. The flowcharts of embedding and extraction processes are given in Fig. \ref{fig:EmbeddingScheme} and Fig. \ref{fig:ExtractingScheme} respectively.
\subsection{Watermark embedding}

In order to ensure both robustness and imperceptibility, the proposed method exploits the full advantages of the mesh saliency and SCS quantification. The watermarking bits are embedded by quantizying the vertex norms of the 3-D model using mesh saliency. The SCS quantization is used since it  is blind and provides a good tradeoff between robustness and capacity \cite{eggers2003scalar}. First, the mesh saliency is computed based on a threshold  fixed automatically to define the salient  points. In fact, for each object, the $70$\% maximum values of saliency vector represent the salient points while the other points are considered non-salient. Second, the norms of salient points are calculated according to this threshold.  The watermark bits are embedded by quantifying the vertex norms of salient points using Lee's mesh saliency \cite{lee2005mesh}.
 Next, the quantization step $Q_S$ is fixed to $N_{av}/\lambda$, where $N_{av}$ is the average of the verices normal norms while $\lambda$  is a parameter which adjusts the tradeoff between robustness  and imperceptibility. This parameter is tuned experimentally and chosen in such a way that ensures  both good robustness and high imperceptibility. The quantization of the salient vertex norms ($NV$) is carried out using the $2$-symbol scalar Costa scheme (SCS) \cite{eggers2003scalar} where a random code is established for each vertex norm using equation  \ref{eq:Codebook}.
\begin{equation}
\label{eq:Codebook}
\beta_{x_{i},t_{x_{i}}}= \bigcup_{l=0}^{1} \left \{ u=zQ_S+l\frac{Q_S}{2}+t_{x_i} \right  \}
\end{equation}

where $z \in \mathbb{Z^+}$, $l \in \left \{ 0,1 \right \}$ represents the watermark bit, $Q_S$ is the quantization step,  $t_{x_i}$ is an additive pseudo-random dither signal generated using a secret key (key1). 
 Next, we look for the nearest codeword $\beta_{NV_i}^J$ to $NV_{i^J}$ in the codebook that implies the correct watermark bit. The quantized value $NV_i^{'J}$  is calculated according to (\ref{eq:Extraction}). The perfect security is achieved when $\gamma =0.5$ as explained in details in \cite{perez2005information}.
 \begin{equation}
 \label{eq:Extraction}
   NV_i^{'J}=\left \| NV_i^J \right \|+\gamma(\beta_{NV_i^J}-\left \| NV_i^J \right \|)   
  \end{equation}
After the quantization process, the dense mesh is  reconstructed starting from the modified vertex norms. We note that the quantization step is used as an extra secret key (key2) which will be used  in the extraction process. For more details, the embedding steps are described in Algorithm \ref{alg:embedding}.

\begin{algorithm}
\caption{Watermark embedding}
\label{alg:embedding}
\begin{algorithmic}
\STATE  1- Compute the saliency of the 3-D original mesh using Lee's method \cite{lee2005mesh} and extract salient vertices.
\STATE  2- Sort in the descending order the  salient vertices  according to their normal norms.
\STATE 3- Calculate the average normal norms of salient vertices $N_{Av}$ of normals and set the  quantization step as $N_{av}/\lambda$. 
\STATE 4- Calculate the norms of the salient vertices and quantize them using $2$-symbol scalar Costa quantization scheme (\ref{eq:Codebook}) according to the predefined order.
\STATE 5- Reconstruct the watermarked dense mesh  starting from the modified vertex norms.
\end{algorithmic}
\end{algorithm}

\subsection{Watermark extraction}
The exaction process is blind since we doesn't need the original mesh. Only the secret keys (key1 and key2) are needed.  Firstly, mesh saliency of the 3-D watermarked mesh is calculated in order to extract  salient vertices according to the same threshold used in the  embedding process. This parameter is chosen automatically since it represents the $70$\% maximum values of the mesh saliency. Secondly, the norms of salient vertices are calculated. Afterwards, we reestablish the vertex order (the norms of vertices normals  sorted in the descending order). Next, we recalculate the quantization step and reconstruct the codebook. Finally, we search the nearest codeword to the vertex  norm in the reconstructed codebook with the aim of finding the watermark bits. For more details, the extraction steps are described in Algorithm \ref{alg:extraction}.

\begin{algorithm}
\caption{Watermark extracting}
\label{alg:extraction}
\begin{algorithmic}
\STATE  1- Calculate the  saliency of 3-D watermarked mesh.
\STATE  2- Extract salient vertices and calculate their corresponding norms.
\STATE  3- Reestablish the vertices order according to the normal norms of salient vertices.
\STATE 4- Recalculate the quantization step and reconstruct the codebook.
\STATE 5- Extract the watermark bits by looking for the nearest codeword to the vertex norms.
\end{algorithmic}
\end{algorithm}


\section{Experimental results}
\label{Experimental results}
\subsection{Experimental setup}
Several experiments were carried out to assess the performance of the  proposed watermarking method on 3-D meshes with different shape complexities: Flower ($2523$ vertices, $4895$ faces), Vase ($2527$ vertices, $5004$ faces), Cup ($9076$ vertices, $18152$ faces), Ant ($7654$ vertices, $15304$ faces), Bimba ($8857$ vertices,  $17710$ faces) and cat ($3534$ vertices, $6975$ faces).  Fig. \ref{fig:OriginalAndWatermarkedObjects}((a),(c),(e),(g)) shows the above mentioned objects. We note  that for comparison purpose the imperceptibility and robustness  have been evaluated using the 3-D meshes: Bunny ($34835$ vertices, $69666$ faces), Horse ($112642$ vertices , $225280$ faces) and Venus ($100759$ vertices, $201514$ faces). The quantification step is chosen in such a way that ensures the best imperceptibility-robustness tradeoff. This parameter has been tuned experimentally and we kept $Q_S=0.08$.

\subsection{Evaluation metrics}
Several experiments were conducted to assess the performance of the proposed method in terms of imperceptibility and robustness. The distortion introduced by the watermark embedding is evaluated  objectively and visually using the maximum root mean square error (MRMS), hausdorff distance (HD) and mesh structural distortion measure (MSDM) respectively. The robustness of the proposed scheme is evaluated using the normalized correlation ($Corr$).
\subsubsection{Imperceptibility}
~\\
In order to evaluate the imperceptibility of the proposed method several  metrics have been used to measure the amount of distortion introduced by the embedding process. This distortion can be measured geometrically or perceptually.
 The maximum root mean square error (MRMS) proposed in \cite{Metro} is used to calculate the objective distortion between the original meshes and the watermarked ones.

 The MRMS which refers to  the maximum between the two root mean square error (RMS) distances calculated by:
 
 \begin{equation}
 d_{MRMS}=max(d_{RMS}(M,Mw),d_{RMS}(Mw,M))
 \end{equation}
\begin{equation}
 d_{RMS}(M,Mw)=\sqrt{\frac{1}{|M|}\int \int_{p\in M} d(p,Mw)^2 dM}
 \end{equation} 
 
 where $p$ is a point on surface $M$, $|M|$ represents the area of $M$, and $d(p,Mw)$ is the point-to-surface distance between $p$ and $Mw$. It is worth noticing that surface-to-surface distance, as the MRMS metric, does not represent the visual distance between the two meshes \cite{lavoue2006perceptually}.   
So, another perceptual metric is needed  to measure the distortion caused by the watermark insertion. 

The mesh structural distortion measure (MSDM) metric is chosen  to measure the visual degradation of the watermarked meshes \cite{lavoue2006perceptually}. The MSDM value is equal $0$ when the original and watermarked 3-D objects are identical . Otherwise, the MSDM value is equal to $1$ when the objects are visually very different.  The global MSDM distance between the original mesh  $M$ and watermarked mesh $Mw$ having $n$ vertices respectively is defined by : 
\begin{equation}
d_{MSDM}(M,M_w)=\left ( \frac{1}{n} \sum_{i=1}^{n} d_{LMSDM}(a_i,b_i)^3 \right )^{\frac{1}{3}} \in [0,1) 
\end{equation}
  $d_{LMSDM}$ is the local MSDM distance  between two mesh local windows $a$ and $b$ (in mesh $M$ and $Mw$ respectively) which is defined by :
 \begin{equation}
 \begin{split}
 d_{LMSDM}(a,b) & = ( 0.4 \times Curv(a,b)^3  + 0.4 \times Cont(a,b)^3  \\   & +  0.2 \times Surf(a,b)^3 )^{\frac{1}{3}} 
\end{split}
\end{equation} 
 $Curv$, $Cont$ and $Surf$ refers to curvature, contrast and structure comparison functions respectively. 

\begin{table}[!h]
\footnotesize
\centering
{\renewcommand{\arraystretch}{1.2}
\caption{Watermark imperceptibility measured in terms of MRMS, HD and MSDM.}
\label{tab:imperceptibilityEvaluation}
\begin{tabular}{|c|c|c|c|c|}
\hline
Model &  \: MRMS ($10^{-3}$) \: & \: HD ($10^{-3}$) \: & \: MSDM   \:  \tabularnewline
 \hline
Flower & $0.60$ & $4.13$ & $0.21$ \tabularnewline
\hline
Vase &$0.38$ & $3.14$ &$0.41$ \tabularnewline
\hline
Cup  & $0.96$&  $2.98$ & $0.42$\tabularnewline
\hline
Ant   & $0.71$ &$4.33$ &$0.57$ \tabularnewline
\hline

 Cat   & $0.67$ & $1.2$ &$0.18$\tabularnewline
\hline
Bimba   & $0.37$ & $1.62$ &$0.11$\tabularnewline
\hline

\end{tabular}}
\\
\end{table}

\begin{figure}[!h]
    \centering
    \subfigure[]{\label{sub21} \includegraphics[width=0.2\textwidth]{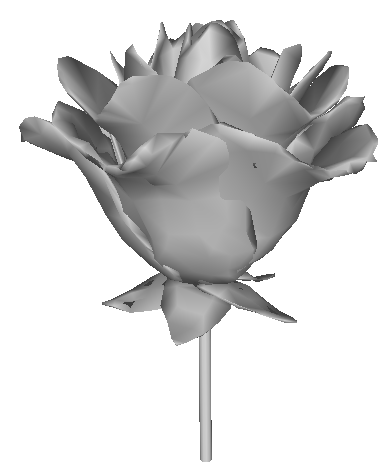}}
    \subfigure[]{\label{sub22} \includegraphics[width=0.2\textwidth]{images/flower.png}}
    \subfigure[]{\label{sub23} \includegraphics[width=0.2\textwidth]{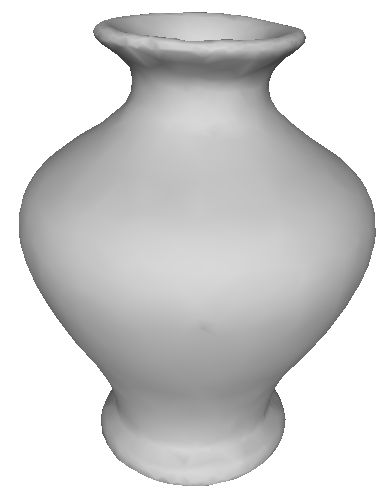}}
    \subfigure[]{\label{sub24} \includegraphics[width=0.2\textwidth]{images/vase.png}}
    \subfigure[]{\label{sub25} \includegraphics[width=0.22\textwidth]{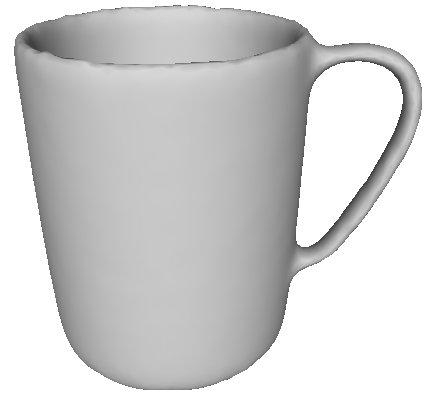}}
    \subfigure[]{\label{sub26} \includegraphics[width=0.22\textwidth]{images/cup.png}}
    \subfigure[]{\label{sub29} \includegraphics[width=0.2\textwidth]{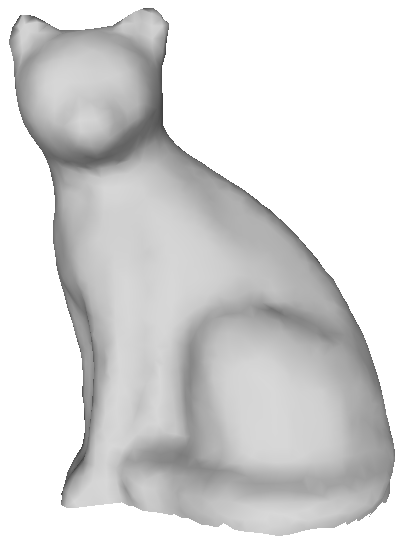}}
    \subfigure[]{\label{sub30} \includegraphics[width=0.2\textwidth]{images/cat.png}}
       \caption{(a) Flower, (b) Watermarked Flower, (c) Vase, (d) Watermarked Vase, (e) Cup, (f) Watermarked Cup, (g) Ant, (h) Watermarked Ant, (i) Cat, (j) Watermarked Cat.}
    \label{fig:OriginalAndWatermarkedObjects}
\end{figure}

\begin{figure}[!h]
    \centering
    \subfigure[]{\label{sub1} \includegraphics[width=0.15\textwidth]{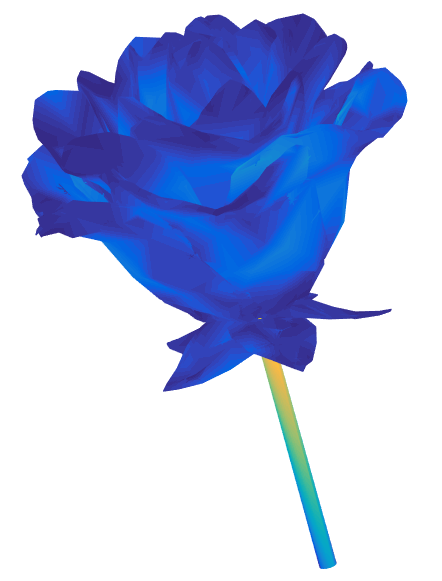}}
    \subfigure[]{\label{sub2} \includegraphics[width=0.15\textwidth]{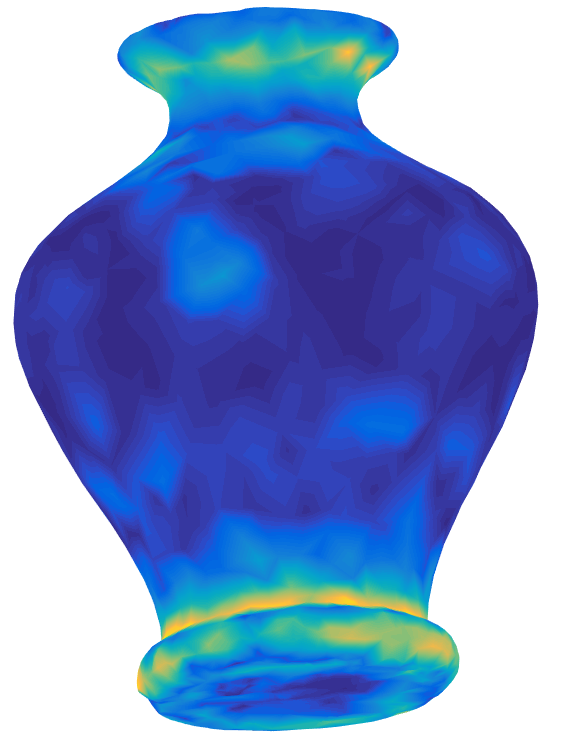}}
    \subfigure[]{\label{sub3} \includegraphics[width=0.15\textwidth]{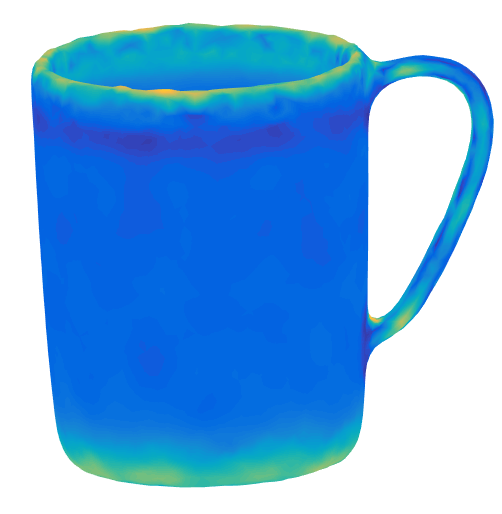}}
    \subfigure[]{\label{sub4} \includegraphics[width=0.15\textwidth]{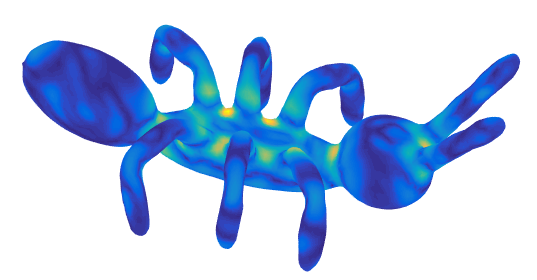}}
    \subfigure[]{\label{sub5} \includegraphics[width=0.15\textwidth]{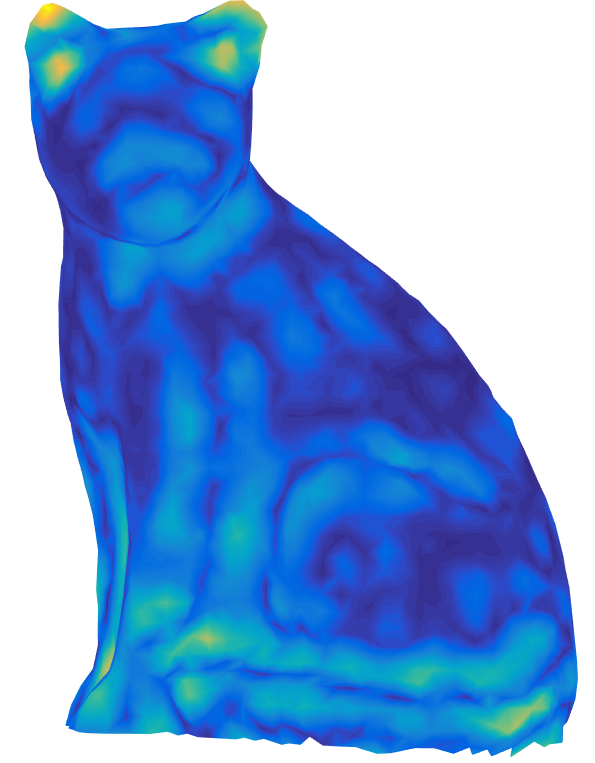}}
     \subfigure[]{\label{sub6} \includegraphics[width=0.15\textwidth]{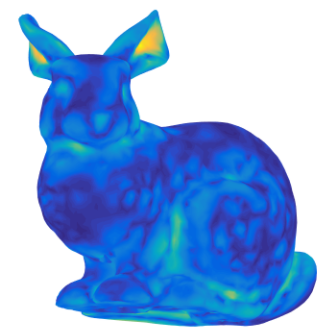}}
       \caption{ Saliency of 3-D meshes using Lee's method \cite{lee2005mesh} (a) Flower, (b) Vase, (c) Cup, (d) Ant, (e) Cat, (f) Bunny.}
    \label{fig:SaliencyObjects}
\end{figure}

\subsubsection{Robustness}
~\\
The robustness is measured using the normalized correlation ($Corr$) between the inserted watermark and the extracted one as given by the following equation : 
\begin{equation}
Corr=\frac {\sum_{i=1}^{m}(w'_i-\overline{w}^*)(w_i-\overline{w})}{\sqrt{\sum_{i=1}^{m}(w'_i-\overline{w}^*)^2.\sum_{i=1}^{m}(w_i-\overline{w})^2}}
\end{equation}
where $i\in \left \{ 1,2,\dots, m \right \}$, $m$ is the length of the watermark, $\overline{w}^*$ and $\overline{w}$ are the averages of the watermark bits respectively.


\subsection{Results and discussion}
\subsubsection{Imperceptibility}
  Fig. \ref{fig:OriginalAndWatermarkedObjects} illustrates the original and watermarked 3-D meshes. We can see that the distortion is very imperceptible. This is due to the saliency adjustment. In addition, according to Table \ref{tab:imperceptibilityEvaluation}, it can be observed that the proposed method can achieve high imperceptibility in terms of MRMS, HD and MSDM. We believe that this performance is obtained thanks to the exploitation of mesh saliency to avoid serious distortions. It can be also observed that the imperceptibility results in terms of MRMS, HD and MSDM are different from a mesh to another. This difference is mainly due to the curvature nature of each one of these 3-D meshes.

\begin{table}[!h]
\footnotesize
\centering
{\renewcommand{\arraystretch}{1.2}
\caption{Watermark imperceptibility without using saliency measured in terms of MRMS, HD and MSDM  compared to the proposed method.}
\label{tab:ImperceptWithoutandWithSaliency}
\begin{tabular}{|c|c|c|c|c|}
\hline
Model &  \: MRMS ($10^{-3}$) \: & \: HD ($10^{-3}$) \: & \: MSDM   \:  \tabularnewline
 \hline
Flower & $0.89/0.60$ & $5.03/4.13$ & $0.88/0.21$ \tabularnewline
\hline
Vase &$0.58/0.38$ & $4.76/3.14$ &$0.76/0.41$ \tabularnewline
\hline
Cup  & $1.02/0.96$&  $3.45/2.98$ & $0.87/0.42$\tabularnewline
\hline
Ant   & $0.83/0.71$ &$4.43/4.33$ &$1.0/0.57$ \tabularnewline
\hline

 Cat   & $1.2/0.67$ & $1.9/1.2$ &$0.29/0.18$\tabularnewline
\hline
Bimba   & $0.76/0.37$ & $2.98/1.62$ &$1.66/0.11$\tabularnewline
\hline

\end{tabular}}
\\
\end{table}

  To further evaluate the importance of using mesh saliency to improve the imperceptibility of the proposed method, we compare the obtained results with those obtained without using the saliency. Table \ref{tab:ImperceptWithoutandWithSaliency} exhibits the imperceptibility performance in terms of MRMS, HD and MSDM without using the mesh saliency compared to the proposed method based on mesh saliency. According to Table \ref{tab:ImperceptWithoutandWithSaliency}, it can be seen that the proposed method gives good results  which illustrates the  imperceptibility improvement achieved using the saliency aspect in the watermark embedding.
\subsubsection{Robustness}
To evaluate the robustness of the proposed scheme, 3-D meshes have been undergone several attacks. For this purpose, a benchmarking system has been used \cite{benchmark2010}. The robustness of our scheme is tested under several attacks including noise addition, smoothing, quantization, cropping, subdivision and similarity transformations (translation, rotation and uniform scaling).  Fig. \ref{fig:AttackedBimba} shows the model Bimba after several attacks.
To evaluate the robustness to noise addition attack, binary random noise was added to each vertex of 3-D models with four different noise amplitudes : $0.05$\%, $0.10$\%, $0.30$\% and $0.50$\%. According to Table \ref{tab:RobustnessNoise}, it can be seen that the proposed method is robust against noise addition four all the 3-D models.

For evaluating the resistance of the proposed scheme to smoothing attack, the 3-D models have undergone Laplacian smoothing proposed in \cite{taubiny2000geometric} using $5$, $10$, $30$ and $50$ iterations while keeping the deformation factor $ \lambda=0.10$. Table \ref{tab:RobustnessSmoothing} shows that our method is able to withstand smoothing operation. The robustness of the proposed scheme is evaluated against elements reordoring attack called also file attack. According to Table \ref{tab:RobustnessElmtReord} the proposed scheme can  resist to element reordoring. Quantization is also applied  to the 3-D models to evaluate the robustness against this attack using  $7$, $8$, $9$, $10$ and $11$ bits. It can be concluded from Table \ref{tab:RobustnessQuantization} that our method shows good robustness  against quantization regardless of the used 3-D mesh. The robustness of the proposed method is evaluated against similarity transformation in which 3-D models have undergone a random rotation,  a random uniform scaling and  a random translation. Table  \ref{tab:RobustnessSimilarTrans} sketches the obtained results in terms of correlation. It can be observed that our method can achieve high robustness against these attacks. Finally, the proposed scheme is tested against subdivision attack including three schemes (loop , midpoint and sqrt3). The obtained results in Table \ref{tab:RobustnessSubdivision} in terms of correlation exhibit the high robustness against subdivision. Cropping  is considered to be one of the most damaging attack since it deletes a region from the 3-D mesh and thus the useful information will be lost. It can be observed from Table \ref{tab:RobustnessCropping} that the proposed method is not enough robust to cropping attacks. In fact, if the deleted surface contains salient points, the extraction process will fail. In the future work, we will search a solution to the issue related to the robustness weakness against this attack.

\begin{table}[!h]
\footnotesize
\centering
{\renewcommand{\arraystretch}{1.2}
\caption{Watermark robustness against additive noise  measured in terms of correlation.}
\label{tab:RobustnessNoise}
\begin{tabular}{|c|c|c|c|c|c|c|}
\hline
Noise intensity & Flower &  Vase  & Cup& Ant & Cat&Bimba\tabularnewline
 \hline
 0.05\%  &$0.97$& $1.0$ &$0.97$&$0.98$&$0.98$&$1.0$\tabularnewline
 0.10\%  &$0.95$& $0.92$ &$0.92$&$0.96$&$0.93$&$0.92$\tabularnewline
 0.30\% &$0.89$& $0.87$ &$0.83$&$0.86$&$0.89$&$0.90$\tabularnewline
 0.50\% &$0.83$& $0.72$ &$0.74$&$0.76$&$0.74$&$0.79$\tabularnewline
\hline
\end{tabular}}
\\
\end{table}

\begin{table}[!h]
\footnotesize
\centering
{\renewcommand{\arraystretch}{1.2}
\caption{Watermark robustness against Laplacian smoothing ($\lambda=0.1$)  measured in terms of correlation.}
\label{tab:RobustnessSmoothing}
\begin{tabular}{|c|c|c|c|c|c|c|}
\hline
Number of iterations & Flower&  Vase&Cup&Ant&Cat&Bimba  \tabularnewline
 \hline
 $5$ & $1.0$ & $0.99$ &$1.0$&$0.99$&$1.0$&$1.0$\tabularnewline
$10$ & $0.99$ & $0.98$ &$0.99$&$0.97$&$0.95$&$0.99$\tabularnewline
$30$ & $0.97$ & $0.96$ &$0.92$&$0.94$&$0.95$&$0.94$\tabularnewline
$50$ & $0.88$ & $0.87$ &$0.89$&$0.85$&$0.90$&$0.90$\tabularnewline
\hline
\end{tabular}}
\\
\end{table}

\begin{table}[!h]
\footnotesize
\centering
{\renewcommand{\arraystretch}{1.2}
\caption{Watermark robustness against elements reordering  measured in terms of correlation.}
\label{tab:RobustnessElmtReord}
\begin{tabular}{|c|c|c|c|c|c|c|}
\hline
Elements reordering&  Flower &  Vase&Cup&Ant&Cat&Bimba  \tabularnewline
 \hline
 Element   &$0.99$& $1.0$ &$0.99$&$1.0$&$1.0$&$0.99$\tabularnewline
reordering $1$ &&&&&& \tabularnewline 
\hline
 Element   &$0.96$& $0.98$ &$1.0$&$0.97$&$1.0$&$0.97$\tabularnewline
reordering $2$ &&&&&& \tabularnewline
\hline 
 Element   &$1.0$& $0.99$ &$0.986$&$0.99$&$0.98$&$0.96$\tabularnewline
reordering $3$ &&&&&& \tabularnewline
\hline
 \end{tabular}}
\\
\end{table}

\begin{table}[!h]
\footnotesize
\centering
{\renewcommand{\arraystretch}{1.2}
\caption{Watermark robustness against  quantization measured in terms of correlation.}
\label{tab:RobustnessQuantization}
\begin{tabular}{|c|c|c|c|c|c|c|}
\hline
Quantization & Flower & Vase &Cup&Ant& Cat&Bimba\tabularnewline
 \hline
 $11$-bits &$1.0$ & $1.0$ &$1.0$&$1.0$&$1.0$&$1.0$\tabularnewline
 $10$-bits&$1.0$   & $0.99$ &$0.99$&$0.99$&$0.98$&$1.0$\tabularnewline
 $9$-bits& $0.99$  & $0.98$ &$0.97$&$0.97$&$0.98$&$0.99$\tabularnewline
 $8$-bits&  $0.93$ & $0.90$ &$0.91$&$0.92$&$0.93$&$0.98$\tabularnewline
 7-bits & $0.80$ & $0.79$ &$0.80$&$0.77$&$0.78$&$0.86$\tabularnewline
\hline
 
 \end{tabular}}
\\
\end{table}

\begin{table}[!h]
\footnotesize
\centering
{\renewcommand{\arraystretch}{1.2}
\caption{Watermark robustness against similarity transformations  measured in terms of correlation.}
\label{tab:RobustnessSimilarTrans}
\begin{tabular}{|c|c|c|c|c|c|c|}
\hline
Similarity  & Flower & Vase&Cup&Ant&Cat&Bimba  \tabularnewline
transformations&&&&&& \tabularnewline
 \hline
  Similarity & $0.99$ &$0.92$ &$1.0$&$0.94$&$1.0$&$0.98$ \tabularnewline
  transformation 1&&&&&& \tabularnewline
  \hline
Similarity  & $0.94$&$0.95$&$1.0$&$0.97$&$1.0$&$0.99$ \tabularnewline
 transformation 2&&&&&& \tabularnewline
 \hline
 Similarity  & $0.97$ &$0.98$&$0.99$&$1.0$& $0.98$&$0.94$ \tabularnewline 
 transformation 3&&&&&&\tabularnewline
\hline
 \end{tabular}}
\\
\end{table}

\begin{table}[!h]
\footnotesize
\centering
{\renewcommand{\arraystretch}{1.2}
\caption{Watermark robustness against similarity transformations  measured in terms of correlation.}
\label{tab:RobustnessSubdivision}
\begin{tabular}{|c|c|c|c|c|c|c|}
\hline
Subdivision  & Flower &Vase&Cup&Ant&Cat&Bimba \tabularnewline
 \hline
  Loop iter $1$  &$1.0$&$0.98$&$1.0$&$0.96$& $0.96$& $0.98$\tabularnewline
 Midpoint iter $1$&$0.94$&$0.88$&$0.84$&$0.92$& $0.93$&$0.95$ \tabularnewline
  Sqrt3 iter $1$  &$0.99$&$0.97$&$0.94$&$1.0$& $0.98$&$0.94$ \tabularnewline
\hline
 \end{tabular}}
\\
\end{table}

\begin{table}[!h]
\footnotesize
\centering
{\renewcommand{\arraystretch}{1.2}
\caption{Watermark robustness against cropping measured in terms of correlation.}
\label{tab:RobustnessCropping}
\begin{tabular}{|c|c|c|c|c|c|c|}
\hline
Cropping  & Flower &Vase&Cup&Ant&Cat&Bimba \tabularnewline
 \hline
  $10$   &$0.53$&$0.57$&$0.65$&$0.63$& $0.59$& $0.52$\tabularnewline
 $30$&$0.46$&$0.34$&$0.39$&$0.46$& $0.34$&$0.27$ \tabularnewline
 $50$  &$0.28$&$0.12$&$0.19$&$0.24$& $0.22$&$0.21$ \tabularnewline
\hline
 \end{tabular}}
\\
\end{table}

\subsection{Comparison with alternative methods}
To further evaluate the performance of the proposed scheme in terms of imperceptibility and robustness we compare it with Cho's \cite{Cho2007},  \cite{wang2008hierarchical}, Zhan's et al. \cite{zhan2014blind}, Rolland-Neviere et al. \cite{rolland2014triangle} and Jeongho Son's et al. \cite{son2017perceptual} schemes. We note that for comparison purpose, we have tested the robustness of our method using the 3-D models Bunny, horse and Venus. 

\begin{figure}[!h]
\centering
\captionsetup{justification=centering}
\includegraphics[width=0.56\textwidth]{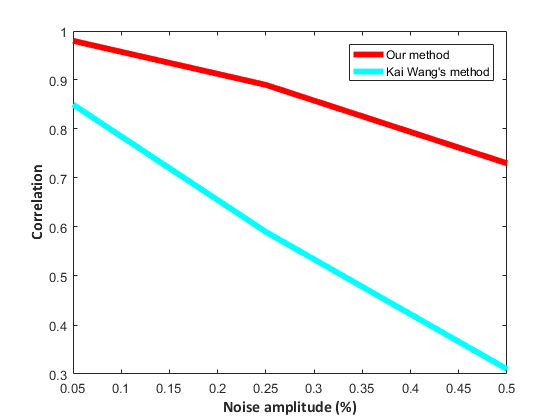}
\caption{Robustness comparison  with Kai Wang's  method \cite{wang2008hierarchical} in terms of correlation against noise addition for Venus model.}
\label{fig:ComparNoiseVenus}
\end{figure}

\begin{figure}[!h]
\centering
\captionsetup{justification=centering}
\includegraphics[width=0.56\textwidth]{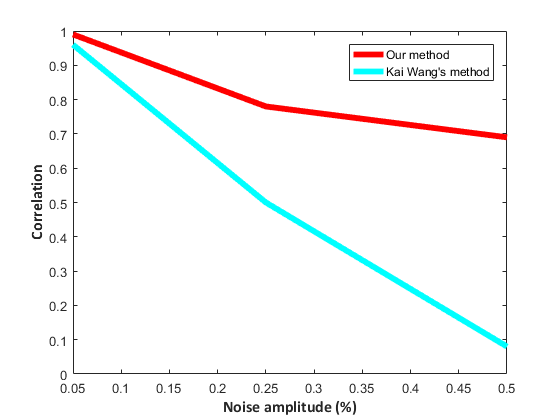}
\caption{Robustness comparison  with Kai Wang's  method \cite{wang2008hierarchical} in terms of correlation against noise addition for Horse model.}
\label{fig:ComparNoiseHorse}
\end{figure}

\begin{figure}[!h]
\centering
\captionsetup{justification=centering}
\includegraphics[width=0.56\textwidth]{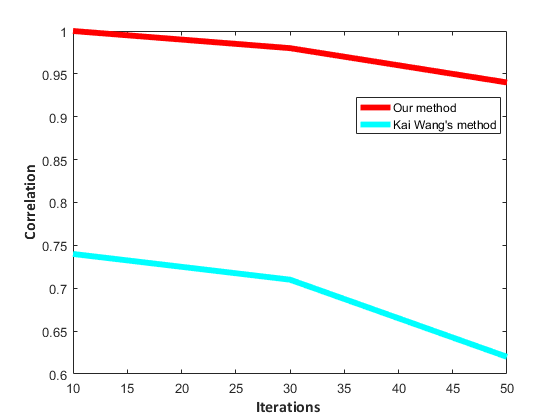}
\caption{Robustness comparison  with Kai Wang's  method \cite{wang2008hierarchical} in terms of correlation against smoothing ($\lambda=0.1 $) for Venus model.}
\label{fig:ComparSmoothVenus}
\end{figure}

\begin{figure}[!h]
\centering
\captionsetup{justification=centering}
\includegraphics[width=0.56\textwidth]{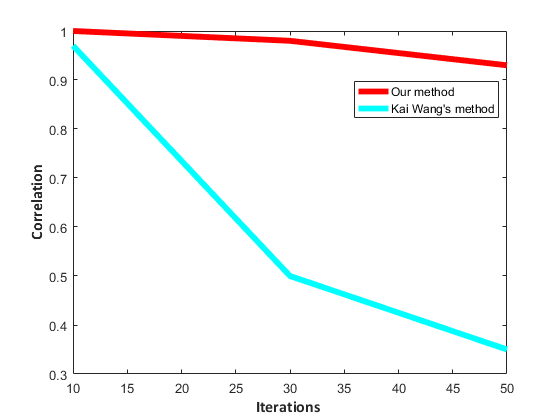}
\caption{Robustness comparison  with Kai Wang's  method \cite{wang2008hierarchical} in terms of correlation against smoothing ($\lambda=0.1 $) for Horse model.}
\label{fig:ComparSmoothHorse}
\end{figure}

\begin{table}[!h]
\footnotesize
\centering
{\renewcommand{\arraystretch}{1.2}
\caption{Imperceptibility comparison with Cho's \cite{Cho2007}, Rolland-Neviere's \cite{rolland2014triangle} and Jeongho Son's \cite{son2017perceptual}   schemes  measured in terms of MRMS and MSDM for Horse model.}
\label{tab:ImperceptibilityComparison}
\begin{tabular}{|c|c|c|}
\hline
Method  & MRMS &MSDM\tabularnewline
&($10^{-3}$)& \tabularnewline
 \hline
  \cite{Cho2007}  &$3.17$&$0.3197$ \tabularnewline
 \cite{rolland2014triangle}&$1.48$&$0.2992$ \tabularnewline
  \cite{son2017perceptual}  &$2.90$&$0.3197$ \tabularnewline
Our method   &$0.51$&$0.2683$ \tabularnewline
\hline
 \end{tabular}}
\\
\end{table}
Table \ref{tab:ImperceptibilityComparison} exhibits the imperceptibility comparison with schemes in terms of MRMS and MSDM. The obtained results demonstrate the high imperceptibility of the proposed method and show its superiority to the alternative methods.
The proposed method is compared to Cho's \cite{Cho2007} and Zhan's \cite{zhan2014blind} methods in terms of imperceptibility in terms of MRMS as well as robustness in terms of correlation against noise addition, smoothing and quantization using Bunny  and Venus 3-D meshes. Tables \ref{tab:NoiseAdditionComparison}, \ref{tab:SmoothingComparison}, \ref{tab:QuantizationComparison} sketch the robustness comparison in terms of correlation between our method and schemes \cite{zhan2014blind} and \cite{Cho2007}. It can be concluded from Tables \ref{tab:NoiseAdditionComparison}, \ref{tab:SmoothingComparison} and \ref{tab:QuantizationComparison} that the proposed method is quite robust to additive noise, smoothing and quantization and outperforms the alternative methods. In addition, according to Figures \ref{fig:ComparNoiseVenus}, \ref{fig:ComparNoiseHorse}, \ref{fig:ComparSmoothVenus}, \ref{fig:ComparSmoothHorse}, it can be seen that the proposed method shows relatively high robustness to Kai Wang's method \cite{wang2008hierarchical} in terms of correlation for noise addition and smoothing.

\begin{table}[!h]
\footnotesize
\centering
{\renewcommand{\arraystretch}{1.2}
\caption{Robustness comparison with Cho's \cite{Cho2007} and \cite{zhan2014blind} schemes  against additive noise in terms of correlation for Bunny and Venus models.}
\label{tab:NoiseAdditionComparison}
\begin{tabular}{|c|c|c|c|c|}
\hline
Model  & Amplitude & \cite{Cho2007} &\cite{zhan2014blind} &Our method\tabularnewline
 \hline
    &$0.1$\%&$0.72$&$1.0$&$1.0$ \tabularnewline
 Bunny&$0.3$\%&$0.72$&$0.91$&$0.94$ \tabularnewline
    &$0.5$\%&$0.66$&$0.80$&$0.86$ \tabularnewline
    \hline
&$0.1$\%&$0.94$&$0.95$&$1.0$ \tabularnewline
 Venus&$0.3$\%&$0.87$&$0.95$& $0.99$\tabularnewline
    &$0.5$\%&$0.27$&$0.79$& $0.83$\tabularnewline
    \hline
 \end{tabular}}
\\
\end{table}

\begin{table}[!h]
\footnotesize
\centering
{\renewcommand{\arraystretch}{1.2}
\caption{Robustness comparison with Cho's \cite{Cho2007} and \cite{zhan2014blind} schemes  against smoothing  in terms of correlation for Bunny and Venus models.}
\label{tab:SmoothingComparison}
\begin{tabular}{|c|c|c|c|c|}
\hline
Model  & Number of & \cite{Cho2007} &\cite{zhan2014blind} &Our method\tabularnewline
 &iterations&&&\tabularnewline
 \hline
    &$10$&$0.84$&$0.92$&$0.96$ \tabularnewline
 Bunny&$30$&$0.60$&$0.85$&$0.92$ \tabularnewline
    &$50$&$0.36$&$0.44$&$0.62$ \tabularnewline
    \hline
&$10$&$0.94$&$0.95$&$0.99$ \tabularnewline
 Venus&$30$&$0.63$&$0.93$& $0.97$\tabularnewline
    &$50$&$0.45$&$0.78$& $0.81$\tabularnewline
    \hline
 \end{tabular}}
\\
\end{table}

\begin{table}[!h]
\footnotesize
\centering
{\renewcommand{\arraystretch}{1.2}
\caption{Robustness comparison with Cho's \cite{Cho2007} and \cite{zhan2014blind} schemes  against quantization  in terms of correlation for Bunny and Venus models.}
\label{tab:QuantizationComparison}
\begin{tabular}{|c|c|c|c|c|}
\hline
Model  & Intensity & \cite{Cho2007} &\cite{zhan2014blind} &Our method\tabularnewline
 \hline
    &$9$&$0.73$&$1.0$&$1.0$ \tabularnewline
 Bunny&$8$&$0.58$&$0.91$&$0.95$ \tabularnewline
    &$7$&$0.17$&$0.58$&$0.65$ \tabularnewline
    \hline
&$9$&$0.87$&$1.0$&$1.0$ \tabularnewline
 Venus&$8$&$0.48$&$0.83$&$0.93$ \tabularnewline
    &$7$&$0.07$&$0.73$&$0.83$ \tabularnewline
    \hline
 \end{tabular}}
\\
\end{table}
%

\section{Conclusion}
\label{Conclusion}
In this work, a blind and robust 3-D mesh watermarking method based on mesh saliency and SCS quantization for Copyright protection is proposed. The proposed method  ensures  both high robustness and  imperceptibility by taking the full advantages of SCS quantization and mesh saliency. The robustness requirement is achieved by quantizing the vertex norms using SCS while the imperceptibility requirement is ensured by adjusting the watermark embedding according to the mesh saliency. The obtained results demonstrate that the proposed scheme yields a good tradeoff between the imperceptibility and robustness requirements. Moreover, experimental results show that in comparison with alternative techniques, the proposed method is able to withstand the majority of common attacks including smoothing, noise addition, quantization, similarity transformations, element reordering, subdivision, etc. Our future work will be focused on  investigating the robustness weakness to cropping attack.

\end{document}